# Modified MBE hardware and techniques and role of gallium purity for attainment of two dimensional electron gas mobility >35 x10$^6$ cm$^2$/Vs in AlGaAs/GaAs quantum wells grown by MBE

Geoffrey C. Gardner[a,b], Saeed Fallahi[a], John D. Watson[a] and Michael J. Manfra[a,b,c,*]

[a]Department of Physics and Astronomy, Birck Nanotechnology Center, and Station Q Purdue, Purdue University, West Lafayette, IN 47907
[b]School of Materials Engineering, Purdue University, West Lafayette, IN 47907
[c]School of Electrical and Computer Engineering, Purdue University, West Lafayette, IN 47907
[*] mmanfra@purdue.edu

**ABSTRACT**

We provide evidence that gallium purity is the primary impediment to attainment of ultra-high mobility in a two-dimensional electron gas (2DEG) in AlGaAs/GaAs heterostructures grown by molecular beam epitaxy (MBE). The purity of gallium can be enhanced dramatically by in-situ high temperature outgassing within an operating MBE. Based on analysis of data from an initial growth campaign in a new MBE system and modifications employed for a 2$^{nd}$ growth campaign, we have produced 2DEGs with low temperature mobility µ in excess of 35x10$^6$cm$^2$/Vs at density n=3.0x10$^{11}$/cm$^2$ and µ=18x10$^6$cm$^2$/Vs at n=1.1x10$^{11}$/cm$^2$. Our 2$^{nd}$ campaign data indicate that gallium purity remains the factor currently limiting µ<40x10$^6$cm$^2$/Vs. We describe strategies to overcome this limitation.

**INTRODUCTION**

The two-dimensional electron gas (2DEG) in AlGaAs/GaAs heterostructures remains a model system for the exploration of strong electron-electron interactions, mesoscale electron transport, and topological order in reduced dimensions [1-16]. The quest to improve AlGaAs/GaAs 2DEG quality is driven by exploration of new phenomena at ever finer energy scales [15]. While several metrics of 2DEG quality including the quantum scattering time ($\tau_q$) derived from low-field Shubnikov-de Haas oscillations, and the size of the excitation gap of fragile fractional quantum Hall effect (FQHE) states ($\Delta_{gap}$) exist, the most commonly employed characterization of quality is low temperature mobility. Mobility defined as µ=e$\tau$/m$^*$, where $\tau$ is the transport lifetime, m$^*$ is the effective mass and e is the magnitude of the electron's charge can be related by the Drude formula to the 2DEG conductivity σ and density n as µ=σ/ne. Mobility is known to be governed at low temperatures (T<1K) by residual background charged impurities in the limit of sufficiently small remote ionized impurity scattering [17-19]. In this regime mobility is a measure of chemical purity of the lattice hosting the 2DEG. Considerable work has been undertaken optimizing heterostructure designs necessary to reach the regime of background impurity limited mobility $>10^7 cm^2/Vs$. Pfeiffer and West at Bell Laboratories [20] demonstrated mobility of $\sim 30 \times 10^6 cm^2/Vs$ with symmetric doping wells above and below the primary quantum well hosting the 2DEG [21]. These results were later confirmed by Umansky and co-workers [13]. However, surprisingly little detail concerning MBE system preparation, MBE growth conditions, source material purity, and source conditioning necessary to produce 2DEGs with $\mu > 30 \times 10^6 cm^2/Vs$ has been presented in the literature. Here we detail the process used in our laboratory to produce samples with mobility greater than $35 \times 10^6 cm^2/Vs$. We present data indicating that gallium purity presently limits mobility in the regime $\mu \leq 40 \times 10^6 cm^2/Vs$, and discuss approaches to overcome this limitation. We anticipate the methods presented here will find utility not only for researchers working with ultra-high quality GaAs but also for those pursuing other quantum materials grown by MBE.

## 2 MBE SYSTEM AND HETEROSTRUCTURE DESIGN

GaAs/AlGaAs heterostructures were grown on 2 in. GaAs (100) substrates by MBE in a modified Veeco Gen II designed specifically to produce ultra-high quality 2DEGs. The main growth chamber is pumped through custom all-metal gate valves and chamber flanges designed to accommodate the full 11.6 in. diameter pump openings of 3 Brooks CT-10 cryopumps which each have a pumping speed of 3000 l/s for air. The CT-10 cryopumps were constructed such that the pump could be baked at >150 °C while operating in order to reduce the outgassing load of the pump's outer vacuum can. An additional titanium sublimation pump was utiliized during the final stage of system preparation. The titanium sublimation pump effectively reduced the hydrogen partial pressure by 1 order of magnitude, from $\sim 10^{-11}$ torr to $10^{-12}$ Torr. The base pressure in the growth chamber after extensive 200°C baking was approximately $1 \times 10^{-12}$ Torr. The background pressure in this regime was completely dominated by hydrogen with all other species exhibiting partial pressures suppressed by at least 2 orders of magnitude as measured with a 200 AMU residual gas analyzer (RGA). The quick-entry load lock and an intermediate outgassing chamber are each pumped by dedicated CT-8 pumps resulting in pressures $1 \times 10^{-9}$ Torr and $<2 \times 10^{-11}$ Torr respectively. All effusion cells, including the large capacity arsenic cell, were custom designed for enhanced thermal efficiency. During growth the gallium source consumes 120 W to produce a GaAs growth rate of 1 monolayer/s while the aluminum cell requires 180 W to produce $Al_{0.24}Ga_{0.76}As$. n-Type doping was accomplished using a home-built silicon filament source that consumes 70 W during deposition and p-type doping was achieved with a home-built carbon filament source that consumes 280 W. The MBE is also equipped with a large area shutter that can be placed between the sources and the substrate manipulator to shield the substrate and manipulator from all source fluxes.

Our version of the now standard "doping well" heterostructure design serves for study of mobility evolution. In these structures the 2DEG resides in a 30 nm quantum well that sits 192 nm below the surface. Si-doping is applied in narrow 2.9 nm GaAs quantum wells surrounded by narrow 2 nm AlAs barriers placed symmetrically 75 nm above and below the principal 30nm quantum well. Fig. 1 shows the simulated [22] band structure of a 30 nm modulation-doped GaAs quantum well with doping wells placed at a 75 nm setback within $Al_{0.24}Ga_{0.76}As$ barriers.

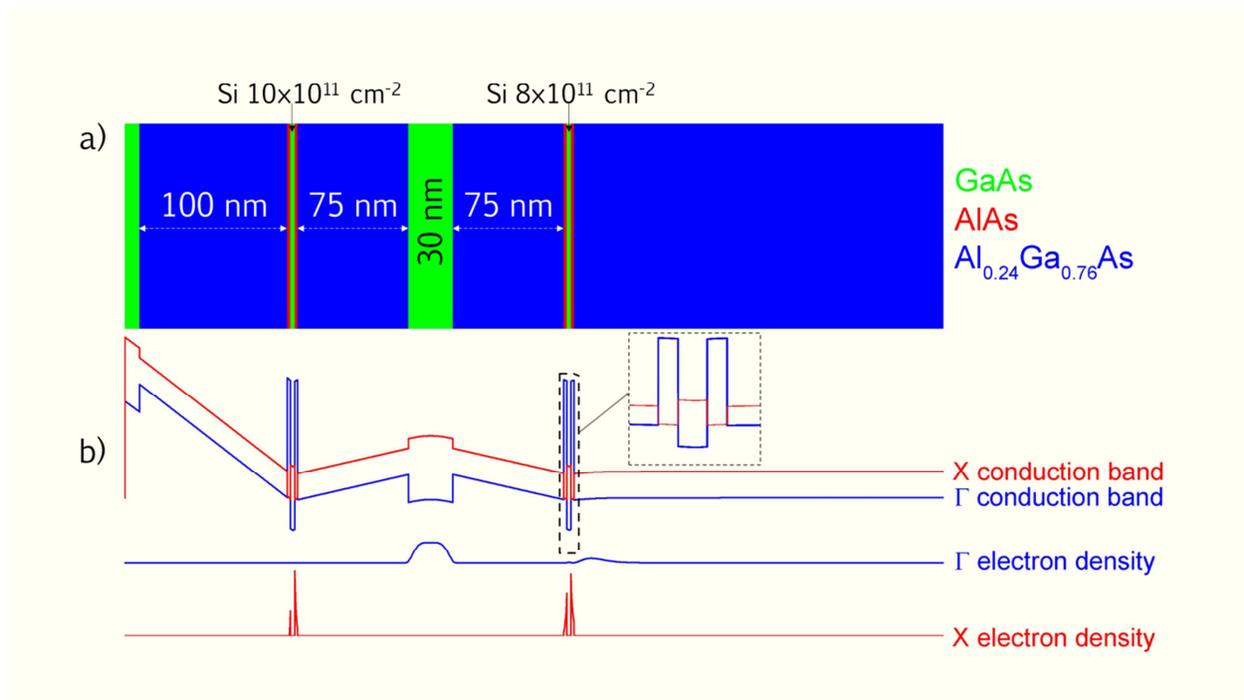

**Fig.1** (a) The layer sequence used to generate a high mobility 2DEG with n=3.0x10$^{11}$cm$^{-2}$.; and (b) $\Gamma$ and X conduction band minimum along the growth direction along with expected location of significant electron density in each band.

The 24% aluminum content of the barrier is sufficient for electron confinement while simultaneously minimizing interface roughness and incorporation of unintentional impurities associated with higher mole fraction barriers. Growth is conducted at $T$=635 °C as measured by optical pyrometry. Growth proceeds at 1µm/hr under the minimal arsenic flux required to prevent surface roughening. In our system this corresponds to a beam-equivalent pressure of 6.5x10$^{-6}$ Torr for arsenic. 20 s growth pauses are introduced on both sides of the principal quantum well to enhance interface smoothness. The temperature of the substrate is decreased to 450°C for the deposition of the silicon atoms to minimize diffusion [23], and a thin layer of GaAs is deposited following the silicon to reduce subsequent surface segregation [24]. The substrate is then rapidly ramped back to 635°C. The doping density is 1x10$^{12}$ atoms/cm$^2$ in the top doping well and 0.8x10$^{12}$ atoms/cm$^2$ the lower doping well. No additional doping is added to the upper barrier for surface compensation. In this structure 3.0x10$^{11}$ electrons/cm$^2$ transfer into the principal 30nm GaAs quantum well. Approximately 6x10$^{11}$ electrons/cm$^2$ transfer to the GaAs surface to compensate the surface states. A significant fraction (approximately 50%) of the electrons remains bound in the X-band of the AlAs barriers flanking the narrow GaAs doping wells. These residual electrons are believed to screen the disorder potential created by the ionized dopants while at the same time not producing a parallel conduction channel in transport measurements due to their low mobility arising from their proximity to their parent silicon ions and their large effective mass in the X band of AlAs.

## 3 RESULTS FROM THE 1st CAMPAIGN

Given the large setbacks used in this symmetrically doped structure to produce electron density $\sim 3\times 10^{11} cm^{-2}$ in a 30 nm well, mobility is primarily limited by unintentional background impurities in the vicinity of the 2DEG [16,18]. Ionized silicon impurity scattering has been shown to be responsible for only ~10% of the total scattering rate [17,19]. The strong dependence of 2DEG mobility on background impurity concentration allows us to study the purity in the GaAs quantum well grown by MBE. Hwang and Das Sarma [18] have calculated that to achieve mobility of $30\times 10^{6} cm^{2}/Vs$ requires a uniform background charged impurity density below $2\times 10^{13} cm^{-3}$.

Our first growth campaign led us to conclude that the purity of gallium charge has a primary impact on 2DEG mobility. The source gallium for the first campaign was 99.99999% (7N) from Alcan [25], which was packaged in high density polyethylene bags. We note that the gallium used in this first campaign, although nominally high purity, was purchased 15 years prior to loading. Newly purchased and packaged 6N5 Aluminum from ULVAC [26] and 7N5 Arsenic from Furukawa [27] were also loaded concomitantly. Despite establishing very good vacuum conditions in a newly constructed MBE system and a reliable heterostructure design, the mobilities measured in the early stages of our 1st growth campaign were extremely poor, $\mu \approx 1\times 10^{5} cm^{2}/Vs$ in simple single heterojunction devices. In order to determine the origin of the poor mobility a series of thick (5-10 micron) unintentionally doped bulk GaAs layers were grown. Electrical characterization of these films revealed an excessively high p-type background carrier concentration, $\sim 1\times 10^{15} cm^{-3}$ in the earliest films. Although we have not yet determined the etiology of this pathology, carbon contamination is a prime suspect. We note that changes to growth conditions including substrate temperature, III/V flux ratio, and growth rate did not impact the residual p-type conductivity, indicating that neither native point defects inherent to the MBE process nor background impurities from the MBE vacuum were a primary source. The MBE vacuum was thoroughly tested and determined to be leak free.

Compromised source material was then considered. In order to determine the precise origin of background we individually subjected the gallium, aluminum, and silicon sources to extended outgassing at high temperature. Given the high vapor pressure of arsenic at temperature above 275 °C, the arsenic cell was not subjected to treatment during this suite of experiments. The gallium sources were first outgassed at 100 °C above their normal growth temperature for 7 h and finally outgassed at 200 °C above growth temperature for 5 h. Outgassing at +200 °C corresponds to a GaAs growth rate of 20μm/h. To accomplish this outgassing in the operating MBE requires some care. The substrate manipulator is first moved to the transfer position such that the wafer puck points away from the sources and the beam flux gauge is aimed at the sources. A large shutter is then placed immediately forward of the ion gauge to protect it and other critical components on the manipulator from the elevated flux. The gallium shutter located immediately ahead of the effusion cell is opened. A small arsenic flux $\sim 2\times 10^{-6}$ Torr is maintained during the cell outgassing to minimize deposition of liquid metal on critical components. Visual inspection through viewports allows us to monitor progress. These outgassing experiments occurred around the 15th growth of the campaign and resulted in an immediate reduction in the p-type background. GaAs films grown after the outgassing experiments were fully depleted for 10 µm thick films. Distinct improvement in 2DEG mobility is seen in Fig. 2. The root of this improvement is believed to be preferential evaporation of high vapor pressure impurities in the gallium. Independent outgassing of the aluminum source resulted in minimal improvement in 2DEG mobility, approximately 10%. As the 2DEG electronic wave function exists primarily in the GaAs quantum well and not in the AlGaAs barrier, this result is reasonable. Outgassing of the silicon source produced no measurable effect. These data clearly indicate that gallium quality was the limiting factor. Most importantly, the data of Fig. 2 indicate that gallium of initially low purity can be purified in the operating MBE by raising the cell temperature well above growth temperature for an extended amount of time. It is not necessary to terminate the campaign and vent the MBE to switch out gallium sources. The data of Fig. 2 also display the typically observed improvement of mobility as the growth campaign progresses. After approximately 75 2DEG growths, the mobility reached $20 \times 10^6$ cm$^2$/Vs, where it saturated (+/- $3 \times 10^6$ cm$^2$/Vs) for the remainder of the campaign.

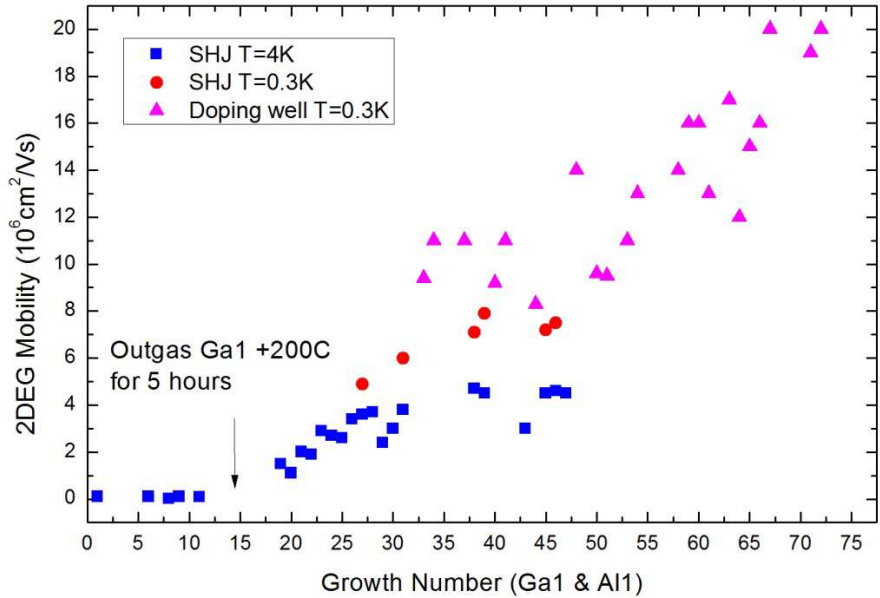

**Fig.2** 2DEG mobility for single interface heterojunction (SHJ) and doping well structures as a function of growth number in the first campaign. An initial period of poor mobility as well as mobility improvement immediately after outgassing gallium is clearly seen. Ga1 refers to the particular cell used.

We have not chemically identified the specific impurities responsible for the initial p-type background and final mobility saturation. Standard chemical analysis techniques lack sensitivity in the regime of interest. Schmult and co-workers [28] discussed the evolution of oxides from a new gallium source as it was initially heated to growth temperature. The stable oxide $Ga_2O_3$ was detected as gaseous $Ga_2O^+$ and $GaH_3O^+$ with a RGA. These signals were strongest during the initial few thermal cycles to growth temperature [28]. The limited sensitivity of the RGA precluded tracking to extremely low partial pressures. It is very likely that surface oxides played an important role in our initial growths as well and that our high temperature outgassing expedited their removal. However oxygen is not expected to incorporate as a shallow acceptor and cannot account for the p-type conductivity observed in our early films [30]. In fact, oxygen may create several distinct traps states within the GaAs bandgap, but none are known to be shallow [30]. It was recently determined that the Alcan gallium from the same lot as used in our 1$^{st}$ campaign also had an abnormally high concentration of germanium. Germanium levels of 440 ppb and 690 ppb were measured by glow discharge mass spectrometry and inductively coupled plasma mass spectrometry respectively [29]. Germanium is amphoteric in GaAs but is likely to incorporate as a p-type impurity under our GaAs growth conditions employed here [30]. Due to its reduced vapor pressure relative to gallium at the temperatures employed in our outgassing experiments, germanium cannot be preferentially evaporated and thus could not be solely responsible for our very low initial mobility and subsequent clean-up. It is possible, however, that germanium was responsible for mobility saturation at $20 \times 10^6$ cm$^2$/Vs.

Despite the inauspicious start to this 1$^{st}$ growth campaign, extremely high quality 2DEGs were ultimately produced [31-35]. As an example, several samples that display the highest activation energies yet measured for the fragile $\nu$=5/2 fractional quantum Hall effect were the product of this campaign. The $\nu$=5/2 state is presently studied for its novel topological properties and potential utility in solid-state quantum computing. In addition samples from the 1$^{st}$ campaign routinely show a strong fractional state at $\nu$=12/5, a state rarely observed and visible only in the highest quality samples [7,36]. Collectively these results demonstrate that an initially poor growth campaign can be salvaged once the source of impurities is identified and corrective measures taken. Our technique relies on the preferential outgassing of high vapor pressure (relative to source material) contaminants.

## 4 MODIFICATIONS TO MBE AND PRACTICES FOR 2$^{nd}$ GROWTH CAMPAIGN

Based upon findings from our first growth campaign, several modifications were made to our gallium charges and MBE hardware in order to decrease the background impurities introduced from source material and to improve the thermal efficiency of system components that produce heat during growth. Improved thermal efficiency of hot components can also potentially reduce background impurity incorporation. As source material handling may also introduce contamination, we also modified our venting and loading procedures. We detail the major changes below.

The substrate heater was redesigned to incorporate increased heater area by increasing the number of resistive heater elements in the assembly from 4 to 6. (see Fig. 3a and b) This 50% increased filament area has the effect of decreasing the temperature of each individual resistive heater element at a given total heater power when compared to our earlier design. The number of layers of heat shielding between the filaments and the remainder of the substrate manipulator was also increased by a factor of 2. Necessary openings in the heat shielding for filament penetrations were reduced from 0.19" to 0.12" in diameter. The opening for the thermocouple penetration was also modified. The total area without direct radiation shielding through which heat could escape decreased by almost 40%. These modifications made the heater assembly more efficient. The power required to reach growth temperature of 635°C dropped from 150 W to 120 W.

Our arsenic cell is a modified Mark V cracker from Veeco Inc [37]. It was redesigned to move the source 1.5" forward, which equates to being more than 16% closer to the substrate. The move forward reduces wasted material and build-up on the cyropanel opening. Two additional heating zones were added to provide better control of the temperature distribution throughout the unit. We added an additional heater between the cracking zone and the arsenic flow valve along the conductance tube in order to remove any potential dead spaces and prevent clogging. We also divided the arsenic reservoir heater into two independent zones and increased the overall length of the heated zone. This modification facilitates finer control over the temperature gradient necessary for reliable and long-term cell operation. As with the substrate heater, additional heat shielding was incorporated to reduce radiative heat loss. The stem valve regulating arsenic flux was newly designed to protect it during material loading and improve reliability by incorporating more a robust valve seat. The net effects of these changes were decreased power consumption, decreased arsenic waste, and decreased arsenic buildup on the cryopanel.

In addition to these MBE hardware modifications we also developed ancillary equipment to aid preparation of new materials and effusion cells to be introduced into the MBE, and to facilitate MBE maintenance without the need for subsequent baking of the system. To minimize the introduction of chemical impurities including water vapor and oxygen during system maintenance, a *mobile* ultra-high vacuum (UHV) chamber was developed (see Fig 3c). The chamber is pumped by a CT-8 cryopump and resides on a rolling stand so that is can be moved into close proximity with the source flange of the MBE. This ultra-clean chamber, which has a pressure below the x-ray limit of a standard ion gauge ($2\times10^{-11}$ Torr), is designed for high temperature outgassing new cells, new crucibles, and evaporative purification of source material. It has 4 source ports and the geometry is designed to allow evaporated gallium to be captured in a vessel at the base of the chamber. It is outfitted with shuttered viewports for optical inspection and temperature measurement. A large shutter protects the pump opening during material evaporation. Diagnostic tools include a 200 AMU RGA. All components introduced to the MBE are first cleaned and characterized in this chamber, and then maintained under UHV conditions until MBE loading.

To minimize impurities introduced during the necessary venting of the MBE a custom glovebox was fabricated. This acrylic enclosure, shown in Fig. 3c, seals to the MBE source flange and the mobile UHV chamber and provides an ultra-pure argon atmosphere, excellent visibility and the necessary manual dexterity required for major MBE maintenance procedures. In our experience, this dedicated glovebox provides significant advantages for major MBE maintenance operations over the commonly employed low density polyethylene glove bags. The glovebox does not develop leaks common to the weak joints of the polyethylene glove bags, and it provides excellent visibility and personal mobility during delicate procedures. Importantly, the large volume and multiple access points for several personnel facilitate complex operations with heavy equipment, e.g. removing and loading the arsenic cell. The volume of the glovebox is actively purged by inflating a large diaphragm internal to the glovebox with pure argon. As this diaphragm expands, it displaces air which exits the glovebox via valved ports. These ports are then closed and the purge argon is released into the glovebox. This process is repeated several times until the appropriate atmosphere is achieved. The environment is analyzed in-situ with a trace oxygen analyzer [38]. The oxygen level drop to below 30 ppm during operation as measured by the trace oxygen analyzer. 99.9999% (6N) research grade argon gas was supplied to both the glovebox and the MBE through orbital welded stainless steel tubes which had been vacuum and heat treated to remove moisture and other impurities. Prior to introduction to the MBE and glovebox the 6N argon gas is passed through a heated titanium gettering furnace, model 2G-100-SS by Centorr [39]. The MBE is also vented to atmospheric pressure with this purified gas. Flow rates into the MBE and glovebox are monitored and controlled through a multiport gas handling manifold that maintains positive flow out of the MBE into the glove box. The glove box is also fitted with a load lock chamber so that tools and materials can be added after the MBE is vented and system is operating under a pure argon environment.

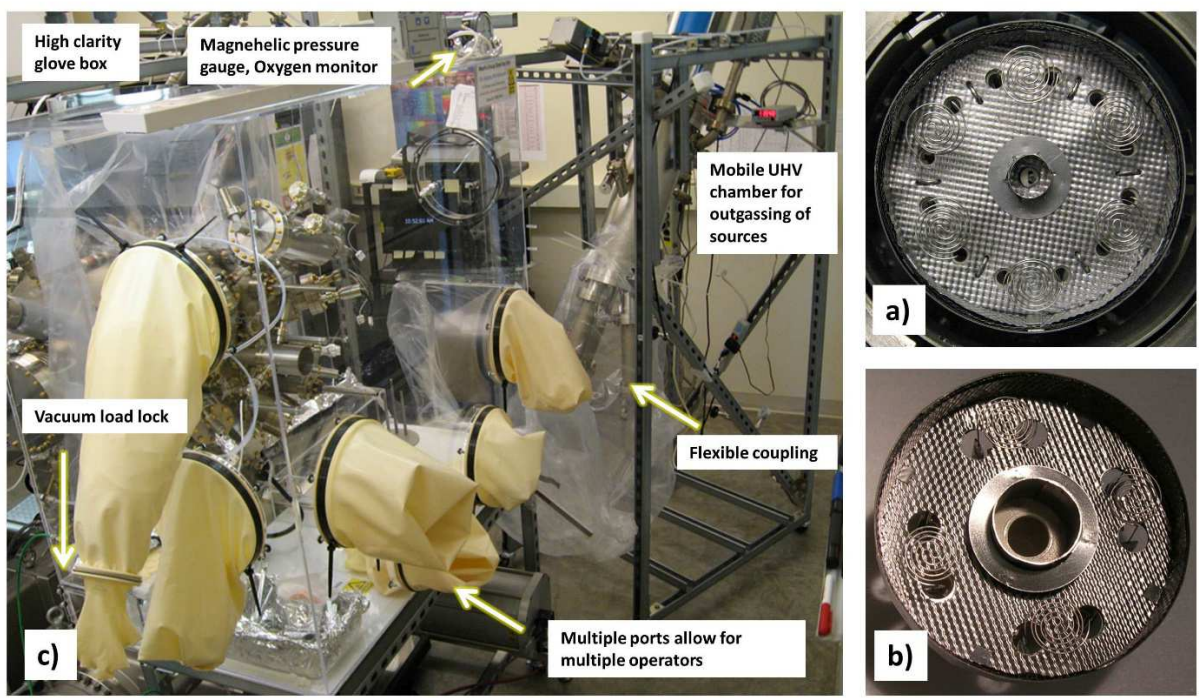

**Fig.3** Clockwise from top right. a) An image of the new 6 filament substrate heater. b) An image of the previous 4 filament substrate heater. The difference in open area for the heat to escape is clear. c) The glove box enclosure seals to the source flange of the MBE chamber as well as a mobile UHV chamber (right), and allows maintenance in an environment with less than 30 ppm of oxygen while still maintaining visual clarity and manual dexterity.

Operation of this system dramatically reduces the introduction of water vapor and oxygen as compared to standard glove bags and obviates the need for 200°C baking after system maintenance. This is particularly benefical in an arsenic-filled MBE which readily absorbs water vapor and oxygen onto large surface areas covered with arsenic. Baking is a time consuming process that may also redistribute contaminants over the entire deposition chamber including, critically, the sources and substrate manipulator. By eliminating baking, the sources and the manipulator stay as contaminant-free as possible.

The efficacy of the improved maintenance techniques, which utilized the glove box mated to mobile chamber and high purity argon delivery, was quantifiable. After warming the MBE to room temperature and immediately prior to venting, the pressure in the MBE was ~$1 \times 10^{-10}$ Torr. After venting in the manner described above, the system returned to below $2 \times 10^{-10}$ Torr within 24 h with little change in the constituent partial pressures. A RGA spectrum is shown in Fig. 4. Of specific concern are compounds containing carbon and oxygen. Oxygen has ability to form a deep trap states while carbon substitutes for the As to serve a shallow acceptor [30]. Peaks to consider in the data are carbon at 12 AMU/e, water vapor at 18 AMU/e, carbon monoxide at 28 AMU/e, oxygen at 32 AMU/e, carbon dioxide at 44 AMU/e, and arsenic oxide at 91 AMU/e. Recall that argon, used as the vent gas, is 40 AMU/e. The most notable increases occur at 18 AMU/e and 40 AMU/e. No substantial increase at 12 AMU/e and 32 AMU/e is observed. Notably absent are hydrocarbon series above 45 AMU/e often seen in recently vented systems. Essentially no new arsenic oxide (91 AMU/e) has been created. 18 AMU/e and 40 AMU/e return to their pre-vent levels within 1 week without any additional action. Helium, 4 AMU/e, was removed by cryopump regeneration. Prior to initiation of our 2$^{nd}$ growth campaign, the MBE was vented approximately 25 times in the manner described above without need for baking.

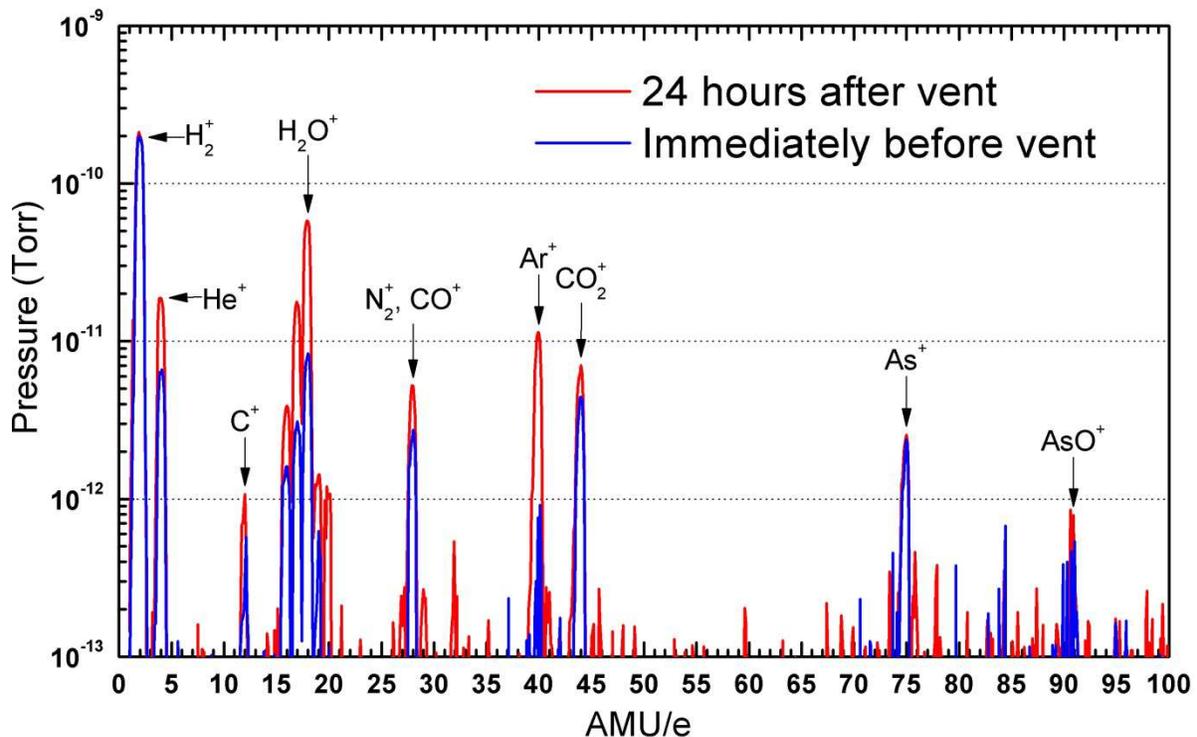

**Fig.4** Characteristic RGA spectra of the vacuum environment 24 h after venting the MBE using techniques detailed in text.

For the 2nd growth campaign two identical effusion cells were loaded with distinct gallium charges. One source was loaded with newly purchased 8N gallium ingots supplied Molycorp Rare Metals Inc. [40]. These ingots were packaged with tantalum sleeves of our own design between the ingot and the plastic packaging to reduce the risk of the plastic contaminating the gallium. At our behest, the Molycorp gallium ingots also underwent an additional proprietary processing step to limit oxide formation [40]. Custom tantalum forceps developed at Purdue were also used for handling the ingots during production and loading. The 2nd gallium effusion cell was loaded with Alcan 7N ingots from the same lot as the first campaign in order to compare source material handled under identical conditions. The arsenic cell was reloaded with material from Furukawa similar to that used in the first campaign. The As ingot was handled with thin tantalum foil rather than being touched directly with gloves. Two aluminum cells were loaded with ULVAC material from the same lot used in the 1st campaign.

## 5 RESULTS OF 2nd CAMPAIGN

As revealed during the first campaign, the purity of the gallium has a primary impact on 2DEG mobility. We conducted several experiments in the 2nd campaign to verify our initial observations, to compare different sources of starting materials, and to verify that mobility evolution is primarily governed the evolution of gallium purity. In the 2nd campaign outgassing experiments were preemptive rather than exploratory. Discrete jumps in mobility are seen in Fig. 5a with each outgassing of the Ga2 cell (Molycorp source material). It is clear that jumps in mobility are unambiguously correlated with each outgassing experiment. Outgassing saves an enormous amount of time. Rather than waiting months to grow several hundred microns of GaAs, the 2DEG mobility can be increased rapidly. Within 10 growths and 2 outgassings the mobility was above $20 \times 10^6 cm^2/Vs$ and rose above $30 \times 10^6 cm^2/Vs$ within 15 runs and 3 outgassings. Similar discrete jumps were also observed for the Ga1 cell (Alcan source material), albeit with lower absolute value of mobility when compared to Ga2. Despite receiving the same number and intensity of high temperature outgassing, Ga2 was consistently able to produce samples of higher mobility than Ga1. The highest mobility achieved with Ga1 again saturated at approximately $20 \times 10^6 cm^2/Vs$. We note that the aluminum cell used (Al1 or Al2) did not influence this conclusion. The differences between Ga1 and Ga2 are most clearly seen in Fig. 5b, where we plot the evolution of mobility of each cell over the course of the campaign. We also confirmed that subsequent and independent outgassing of the aluminum cells had only minor impact on 2DEG mobility, consistent with results from the 1st campaign. We would like to emphasize that the benefits of this in-situ outgassing procedure must be weighed against possible inadvertent damage to the MBE and the need to continue to grow useful structures for experiments. The substrate manipulator, of course, has moving parts that can be damaged by excessive material buildup and one must be cognizant that the gallium movement does not damage the crucible and effusion cell during the high temperature operations.

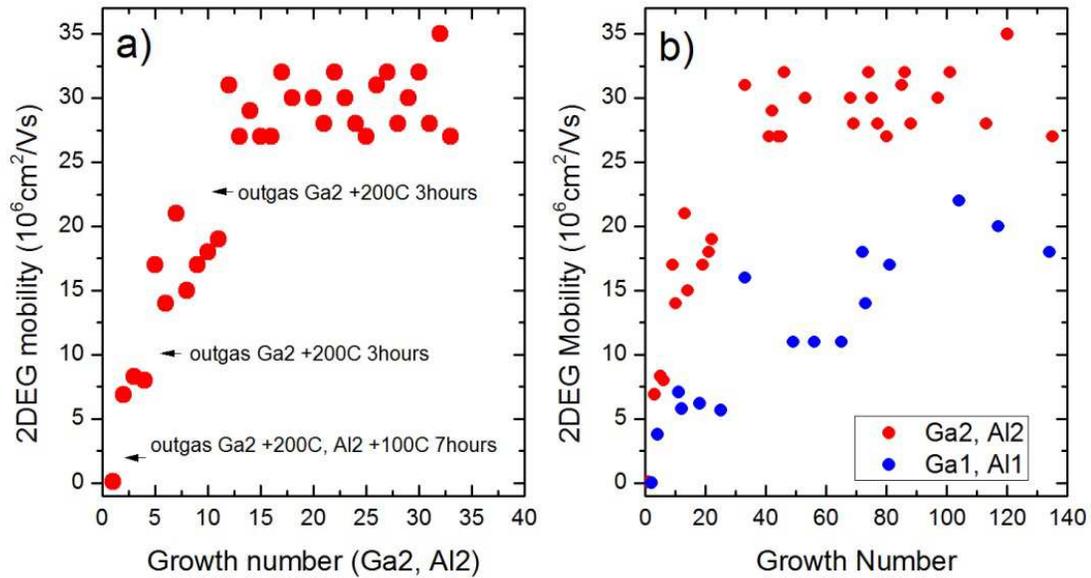

**Fig.5** a) Discrete jumps in mobility are demonstrated using our high temperature outgassing procedure. b) Plot of 2DEG mobility as a function of growth number demonstrates the difference in improvements between the two different gallium source materials despite receiving the same treatment.

Magnetotransport measurements were used to characterize grown wafers at low temperatures and high magnetic field. Electron density and mobility in the quantum well are measured in the Van der Pauw configuration on 4mm x 4mm squares at T=0.3K using standard lock-in techniques. Ohmic contacts were made by annealing eight small InSn blobs around the perimeter at 450 °C for 15 min in forming gas. Fig. 6 shows the measurement results for high 2DEG density and low 2DEG density samples. Electron mobility exceeding $35 \times 10^6$ cm$^2$/Vs is achieved at n=$3.0 \times 10^{11}$ cm$^{-2}$ with well-developed fractional quantum Hall states. This data suggests that the background charged impurity density has been reduced to ~$1 \times 10^{13}$cm$^{-3}$[18]. Particularly noteworthy is the development of the fractional quantum Hall series converging to ν=1/2 in the low density sample. The appearance of many higher order fractions (e.g. up to ν=11/21 at T=0.3K) is an indication not only high mobility but also of high 2DEG density uniformity and proper screening of residual potential fluctuations caused by the remote ionized donors.

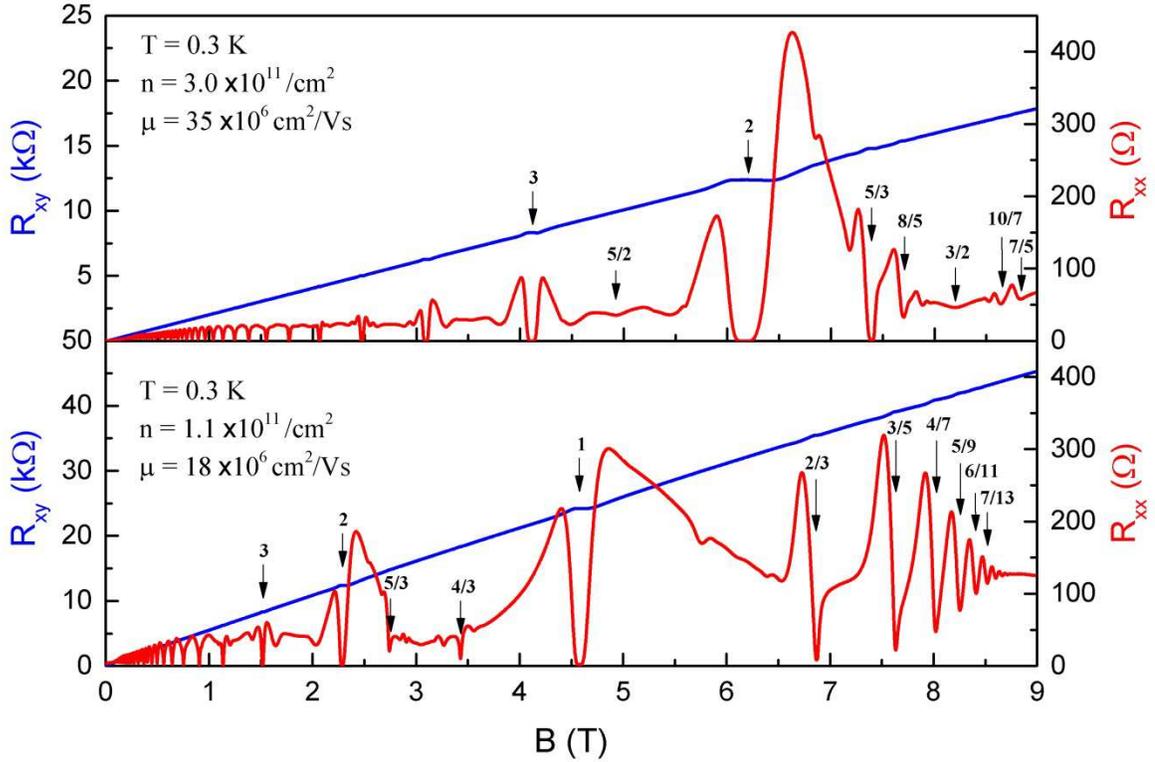

**Fig.6** The top panel shows transport data from a sample with mobility of 35×10$^6$ cm$^2$/Vs and a density of 3.0×10$^{11}$ cm$^{-2}$. The bottom panel shows transport data from a sample with lower density of 1.1×10$^{11}$/cm$^{-2}$ and mobility of 18×10$^6$ cm$^2$ /V s displaying strong fractional state development near ν=1/2.

## 6 CONCLUSION

Modifications employed for our 2$^{nd}$ campaign have yielded a 75% increase mobility when compared to our prior results. More importantly, these results highlight that gallium purity is the critical parameter governing the current state-of-the-art of AlGaAs/GaAs 2DEGs. These findings pave the path to further improvement. As the field is currently limited by the quality of commercially available gallium, individual research groups now need to improve it. We have constructed a laboratory-scale zone-refining apparatus to further purify commercially available gallium in-house. This system is currently in the testing phase and will be the subject of a future publication. We believe that the combination of in-house zone-refining, UHV evaporative purification, and the MBE maintenance and loading practices described here hold promise for future gains. We expect lessons learned in the AlGaAs/GaAs material system will be applicable to other quantum materials grown by MBE.

**ACKNOWLEDGEMENTS**

This work was supported by the W.M. Keck Foundation (Grant no. 52445). J.D.W. was supported by the U.S DOE Office of Basic Energy Sciences, Division of Materials Sciences and Engineering Award no. DE-SC00006671. Addition support from the Office of Naval Research, ONR Grant no. 106378, and Microsoft Research Station Q is gratefully acknowledged. G.C.G. thanks S. Sprowl for helpful discussion regarding glove box fabrication.

**REFERENCES**

[1]  D.C. Tsui, H.L. Stormer, and A.C. Gossard,  Phys. Rev. Lett. **48,** 1559 (1982)

[2] R. de-Picciotto, M. Reznikov, M. Heiblum, V. Umansky, G. Bunin and D. Mahalu, Nature **389**, 162 (1997)

[3] I. P. Radu, J. B. Miller, C. M. Marcus, M. A. Kastner, L. N. Pfeiffer, and K. W. West, Science **320**, 899 (2008)

[4] D. T. McClure, W. Chang, C. M. Marcus, L. N. Pfeiffer, and K. W. West, Phys. Rev. Lett. **108**, 256804 (2012)

[5] R. L. Willett, L. N. Pfeiffer, and K. W. West, Proc. Natl Acad. Sci. **106**, 8853 (2008)

[6] R. L. Willett, C. Nayak, K. Shtengel, L. N. Pfeiffer and K. W. West, Phys. Rev. Lett. **111**, 186401 (2013)

[7] A. Kumar, G. A. Csathy, M. J. Manfra, L. N. Pfeiffer, and K. W. West, Phys. Rev. Lett. **105**, 246898 (2010)

[8]  N. Samkharadze, K. A. Schreiber, G. C. Gardner, M. J. Manfra, E. Fradkin, and G. A. Csathy, Nat. Phys. 3523 (2015) DOI: 10.1038/NPHYS3523

[9]  J.R. Petta, A.C. Johnson, J.M. Taylor, E.A. Laird, A. Yacoby, M.D. Lukin, C. M. Marcus, M.P. Hanson, A.C. Gossard, Science **309,** 2180 (2005).

[10]  M. D. Shulman, S. P. Harvey, J. M. Nichol, S. D. Bartlett, A. C. Doherty, V. Umansky, and A. Yacoby, Nat. Comm. **5**, 5156 (2014)

[11] S. Das Sarma, M. Freedman, and C. Nayak, Phys. Rev. Lett. **94**, 166802 (2005)

[12] C. Nayak, S. H. Simon, A. Stern, M. Freedman, and S. Das Sarma, Rev. Mod. Phys. **80**, 1083 (2008)

[13] V. Umansky, M. Heiblum, Y. Levinson,  J. Smet, J. Nübler, M. Dolev, J. Cryst. Growth **311**, 1658-1661 (2009)

[14] G. Gamez and K. Muraki, Phys. Rev. B **88**, 075308 (2013)

[15] M. J. Manfra, Annu. Rev. Condens. Matter Phys. **5,** 347–73 (2014)

[16] S. Das Sarma, and E. H. Hwang, Phys. Rev. B **90**, 035425 (2014)

[17] L. N. Pfeiffer, K.W. West, H.L. Stormer, and K.W. Baldwin, Appl. Phys. Lett. **55**, 1888 (1989)

[18] E. H. Hwang, and S. Das Sarma, Phys. Rev. B **77**, 235437 (2008)

[19] V. Umansky, R. de-Picciotto, and M. Heiblum, Appl. Phys. Lett. **71**, 683 (1997)

[20] L. N. Pfeiffer, and K. W. West, Physica E **20**, 57-64 (2003)

[21] K. Friedland, R. Hey, H. Kostial, R. Klann, and K. Ploog, Phys. Rev. Lett. **77**, 4616-4619 (1996)

[22] Nextnano3 simulator, © 2012–2015 Walter Schottky Institute, http://www.nextnano.de/nextnano3/index.htm

[23] A. P. Mills Jr., L. N. Pfeiffer, K. W. West, and C. W. Magee, J. Appl. Phys. **88**, 4056 (2000)

[24] L. N. Pfeiffer et. al., Appl. Phys. Lett. **58**, 2258 (1991)

[25] Rio Tinto Alcan, 1188 Sherbrooke Street West, Montreal, Quebec, H3A 3G2, Canada

[26] ULVAC Technologies Inc., 401 Griffin Brook Drive, Methuen MA, 01844

[27] FURUKAWA CO. LTD., 2-3 Marunouchi 2-Chome, Chiyoda-ku, Tokyo 100-8370, Japan

[28] S. Schmult, S. Taylor, and W. Dietsche, J. Cryst. Growth **311**, 1655–57 (2009)

[29] Private communication with Kyungjean Min, David Johnson and Kevin Trumble (Purdue Univ.)

[30] E. F. Schubert, Doping in III–V Semiconductors, Cambridge Univ. Press (1995)

[31] N. Samkharadze, J. D. Watson, G. C. Gardner, M. J. Manfra, L. N. Pfeiffer, K. W. West, and G. A. Csáthy, Phys. Rev. B **84**, 121305 (2011)

[32] N. Deng, G. C. Gardner, S. Mondal, E. Kleinbaum, M. J. Manfra, and G. A. Csáthy, Phys. Rev. Lett. **112**, 116804 (2014)

[33] J. D. Watson, G. A. Csáthy, and M. J. Manfra, Physical Review Applied **3**, 064004 (2015)

[34] M. Samani, A. V. Rossokhaty, E. Sajadi, S. Lüscher, J. A. Folk, J. D. Watson, G. C. Gardner, and M. J. Manfra, Phys. Rev. B **90**, 121405(R) (2014)

[35] N. Deng, J. D. Watson, L. P. Rokhinson, M. J. Manfra, and G. A. Csáthy, Phys. Rev. B **86**, 201301 (2012)

[36] W. Pan, N. Masuhara, N.S. Sullivan, K.W. Baldwin, K.W. West, L.N. Pfeiffer, D.C. Tsui, Phys. Rev. Lett. **106**, 206806 (2011)

[37] Veeco MBE Systems, 4875 Constellation Drive, St. Paul, MN 55127

[38] Alpha Omega Instruments Corp., 40 Albion Road, Suite 100, Lincoln, RI 02865

[39] Centorr Vacuum Industries, 55 Northeastern Blvd, Nashua, NH 03062

[40] Molycorp Inc., 5619 DTC Parkway Suite 1000, Greenwood Village, CO 80111